\documentclass[11pt,twoside]{article}
\usepackage{./asp2014}
\aspSuppressVolSlug
\resetcounters
\usepackage{graphicx}
\newcommand{\ro}{\,$R_{\sun}$}
\newcommand{\cmt}{\,cm$^{-3}$}
\begin{document}

\title{On the diversity and similarity of outbursts of symbiotic 
       binaries and cataclysmic variables}
\author{Augustin Skopal
\affil{Astronomical Institute of the Slovak Academy of Sciences, 
       Tatransk\'a Lomnica, Slovakia;\email{skopal@ta3.sk}}}

\paperauthor{Augustin~Skopal}{skopal@ta3.sk}{ORCID_Or_Blank}
{Astronomical Institute}{Slovak Academy of Sciences}
{Tatransk\'a Lomnica}{}{062 01}{Slovakia}

\begin{abstract}
Outbursts in two classes of interacting binary systems,
the symbiotic stars (SSs) and the cataclysmic variables (CVs),
show a number of similarities in spite of very different
orbital periods. Typical values for SSs are in the order of 
years, whereas for CVs they are of a few hours. 
Both systems undergo unpredictable outbursts, characterized 
by a brightening in the optical by 1--3 and 7--15\,mag for 
SSs and CVs, respectively. 
By modelling the multiwavelength SED of selected examples from 
both groups of these interacting binaries, I determine their 
basic physical parameters at a given time of the outburst 
evolution. In this way I show that the principal difference 
between outbursts of these objects is their violence, whereas 
the ionization structure of their ejecta is basically very 
similar. This suggests that the mechanism of the mass 
ejection by the white dwarfs in these systems is also similar. 
\end{abstract}

\section{Introduction}
Symbiotic stars (SSs) and cataclysmic variables (CVs) are interacting 
binary systems, in which the accretor is a white dwarf (WD). In the former 
the donor is a red giant, while in the latter it is, in most cases, 
a red dwarf. Orbital periods are extremely different, being typically 
in the order of years for SSs, but only of a few hours for CVs. 
The red giant in SSs underfills its Roche lobe with a factor of 
$\sim 0.5$ \citep[e.g.][]{m+s99}, whereas the evolved dwarf donor 
in CVs fills its Roche lobe. This difference dictates 
the way of the principal interaction between the binary components. 
The WD in SSs accretes the matter from the stellar wind of the cool 
giant, while in CVs the mass is transferred onto the WD through 
the $L_1$ point. A review of CVs can be found in the monograph of 
\cite{warner95}, and that on SSs in \cite{siv+mun12}. 

A common feature of these types of interacting binaries are their 
unpredictable outbursts observed on a very different and variable 
time-scale. Here, we compare classical nova (CN) outbursts of CVs 
and Z~And-type outbursts of SSs. 
The former are characterized by a large brightness amplitude 
of $\sim 7-15$\,mag, whereas the latter are as low as 
$\sim 1-3$\,mag in the optical. CN outbursts are believed to be 
caused by a thermonuclear runaway event on the WD surface, when 
the accreted matter exerts the critical pressure, at which 
hydrogen ignites a thermonuclear (CNO) fusion. The CN events 
need accretion onto the WD at rates 
$\dot M_{\rm acc} \lesssim 10^{-8}$\,$M_{\odot}\,{\rm yr}^{-1}$, 
giving the recurrence time of CN explosions much longer than 
human timescales \citep[][]{yaron+05}. A recent review on CNe 
is provided by \cite{bo+ev08}. 
As concerns to the Z~And-type outbursts, their nature is not well 
understood yet. To explain the high luminosity of hot components 
in SSs, it was suggested that they are powered by stable hydrogen 
nuclear burning on the WD surface, which requires accretion rates 
of $\sim 10^{-8} - 10^{-7}$\,$M_{\odot}\,{\rm yr}^{-1}$, depending 
on the WD mass \citep[][]{paczyt78}. The outbursts could 
result from an increase in the accretion rate above that sustaining 
the stable burning, which leads to expansion of the burning 
envelope to an A--F type pseudophotosphere \citep[e.g.][]{tutyan76}, 
and causes a brightening in the optical. 
Multiwavelength modelling of the SED during Z~And-type outbursts 
revealed that the warm pseudophotosphere is simulated by the outer 
flared rim of an edge-on disk around the WD with the nebula 
located above/below the disk \citep[][]{sk05}. Applying the method 
to the explosion of CN V339~Del (Nova Delphini 2013) 
a similar ionization structure of the ejecta was indicated 
\citep[see][]{sk+14}. 

In this contribution I suggest an idea that, in spite of 
an extreme difference in the energetic output of a CN and 
Z~And-type outburst, basic ionization structure, and thus 
the mechanism of the mass ejection by the WD in SSs and CVs 
is similar. 

\section{Multiwavelength modelling of the SED}

The observed continuum spectrum, $F(\lambda)$, emitted by a CN 
in the broad range from the supersoft X-rays to the radio can 
be approximated by two main components: (i) a stellar component, 
$\mathcal{F}_{\lambda}(T_{\rm eff})$ that is produced by the 
WD pseudophotosphere, and (ii) a nebular component represented 
by the volume emission coefficient, 
$\varepsilon_{\lambda}(T_{\rm e})$, of the nebular continuum 
generated in thermal plasma of the ejecta. Therefore, the CN 
spectrum can be expressed as a superposition of these 
components of radiation, i.e.,
%
% -------------------- Eq. (1) ------------------------
%
\begin{equation}
 F(\lambda) = (\theta_{\rm WD}^{\rm eff})^2\,
               \mathcal{F}_{\lambda}(T_{\rm eff}) +
               k_{\rm n} \varepsilon_{\lambda}(T_{\rm e}). 
\label{eq:1}
\end{equation}
The stellar component can be compared with a synthetic spectrum 
calculated for the effective temperature, $T_{\rm eff}$. 
The model is scaled to the observed fluxes by the angular 
radius 
$\theta_{\rm WD}^{\rm eff} = R_{\rm WD}^{\rm eff}/d$, 
given by the effective radius of the pseudophotosphere, 
$R_{\rm WD}^{\rm eff}$ and the distance $d$. 
The factor $k_{\rm n}$ scales the nebular contribution 
$\varepsilon_{\lambda}(T_{\rm e})$ to observations. 
Furthermore, the electron temperature, $T_{\rm e}$, and thus 
$\varepsilon_{\lambda}(T_{\rm e})$ are assumed to be constant 
throughout the nebula, an assumption which simplifies 
determination of its emission measure \textsl{EM}. 
For $T_{\rm eff} \gtrsim 15000$\,K, the radiation of the WD 
pseudophotosphere in the optical can be approximated by the
radiation of a blackbody. 
Fitting parameters are $\theta_{\rm WD}$ and $T_{\rm eff}$, 
which define $R_{\rm WD}^{\rm eff}$ and the luminosity 
$L_{\rm WD} = 4\pi d^2 \theta_{\rm WD}^2 T_{\rm eff}^4$, and 
$k_{\rm n}$ and $T_{\rm e}$, which define 
\textsl{EM} = $4\pi d^2 k_{\rm n}$. 
In the case of modelling the SED of a SS, the component of 
radiation from the giant, 
$\theta_{\rm g}^2 \mathcal{F}_{\lambda}(T_{\rm eff,g})$, 
is added to the model \citep[see][ in detail]{sk05}. 
%
%-------------------------- Fig. 1 ------------------------
%
\begin{figure*}[!t]
%\centering    
\begin{center}
\resizebox{\hsize}{!}{\includegraphics[angle=-90]{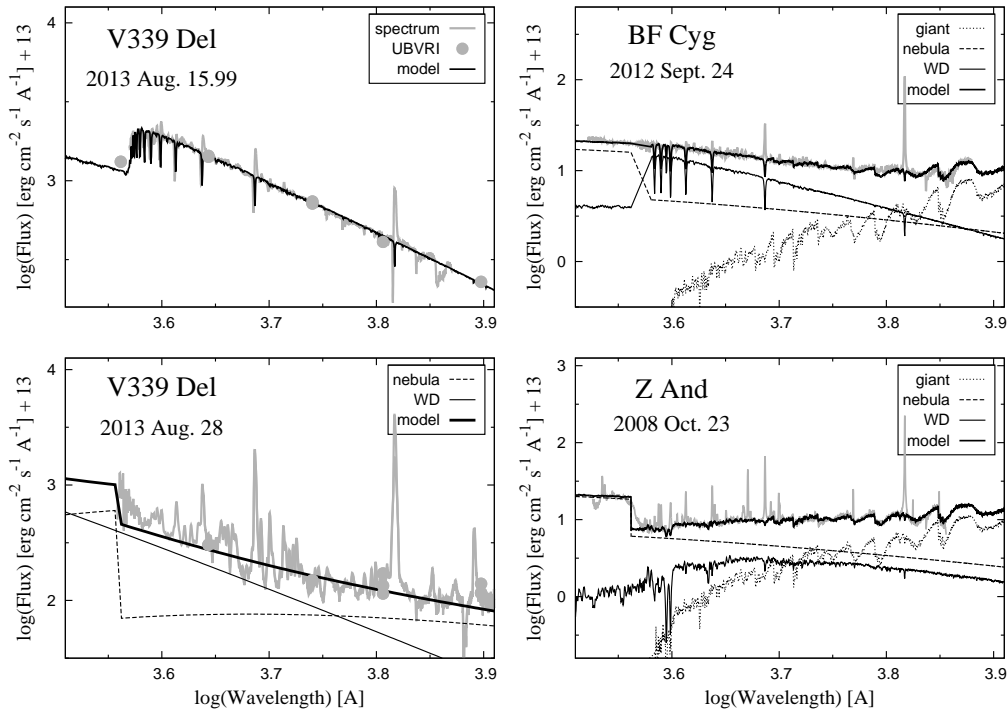}}
\end{center}
\caption[]{
Examples of the optical spectra (gray lines) and their models 
(heavy solid lines).
{\em Left:} The CN V339~Del \citep{sk+14}. 
{\em Right:} The SSs BF~Cyg and Z~And during active phases
\citep{tarsko12}.
Physical parameters of individual radiation components
are collected in Table~1. 
          }
\label{fig:seds}
\end{figure*} 

\section{Example objects}

\subsection{Classical nova V339~Del (Nova Delphini 2013)}

Nova Delphini 2013 (V339~Del) was discovered by Koichi Itagaki 
on 2013 Aug.~14.584 UT at the unfiltered brightness of 6.8 mag 
(CBET No.~3628). Its brightness peaked at $V\sim 4.43$ on 
Aug. 16.44 UT, i.e. $\sim 1.85$ days after its discovery. 
First detailed description of the multicolour optical photometry
was provided by \cite{munari+13} and \cite{chochol+14}, who 
classified the object as a fast nova, placed at a distance 
of 3\,kpc with a low reddening on the line of sight, 
$E_{\rm B-V}$ = 0.18\,mag. A detailed analysis of the nova 
evolution from its discovery to day $\sim 40$ after its optical 
maximum was provided by \cite{sk+14}. 
During the fireball stage (Aug. 14.8--19.9, 2013), $T_{\rm eff}$ 
of the WD pseudophotosphere was in the range of 6000--12000\,K, 
$R_{\rm WD}^{\rm eff}$ was expanding from 
$\sim$\,66 to $\sim$\,300\,$(d/3\,{\rm kpc})$\ro\ and $L_{\rm WD}$ 
was super-Eddington, but not constant. Contribution to the nebular 
continuum was negligible. 
After the fireball stage, a large \textsl{EM} of 
$1.0-2.0\times 10^{62}\,(d/3\,{\rm kpc})^2$\cmt\ was present 
in the spectrum of nova V339~Del. Examples of models SED 
from the fireball stage and just after it are shown in 
Fig.~\ref{fig:seds}. 

\subsection{Symbiotic star BF~Cyg}

BF~Cyg is an eclipsing symbiotic binary with an orbital period  
of 757.2\,d, whose donor is a late-type M5\,III giant 
\citep[e.g.][]{fekel+01}. Its light curve (LC) 
occasionally shows outbursts of the Z~And-type (1920, 1989, 
2006). During the 2006 August, BF~Cyg underwent an outburst 
which continues to the present. Spectroscopic observations 
showed strong P-Cyg type of H{\small I}, He{\small II} lines 
from the beginning of the outburst \citep[e.g.][]{mckeever+11}. 
Recently, \cite{skopal+13} reported an evidence of 
highly-collimated bipolar ejection from the system. 
The model SED from the recent 2006 outburst is shown 
in Fig.~\ref{fig:seds}. 
%
%-------------------------------
%    Table 1: SED's parameters  
%-------------------------------
%
\begin{table}[!t]
\begin{center}
\scriptsize
%\small
\caption{Physical parameters of the CN V339~Del and symbiotic binaries 
         BF~Cyg and Z~And derived from their models SED (Fig.~1). 
         $L_{\rm ph}$ is the rate of hydrogen-ionizing photons 
         required to produce the measured \textsl{EM}. 
}
\begin{tabular}{cccccccccc}\\[-2mm]
\hline
Object                 &
Date                   &
$d$                    &
$T_{\rm eff}$          &
$R_{\rm WD}^{\rm eff}$ &
$L_{\rm WD}$$^{1)}$    &
$T_{\rm e}$            &
\textsl{EM}$^{2)}$     &
$L_{\rm ph}$           &
$\dot M_{\rm WD}$      \\
                       &
dd/mm/yyyy             &
[kpc]                  &
[K]                    &
[$R_{\odot}$]          &
                       &
[K]                    &
                       &
[s$^{-1}$]             &
[$M_{\odot}yr^{-1}$]   \\
\hline
V339~Del$^{3)}$ & 16/08/2013 & 3.0 & 12000 & 110 &  86&   --   &   --
         &         --               &  --    \\
         & 28/08/2013 & 3.0 & 43000 &  13 & 210& 15000  &  180
         &  3.2$\times 10^{49}$ & 4.5$\times 10^{-4}$ \\
BF~Cyg$^{4)}$   & 24/09/2012 & 3.8 &  8750 &  19 &0.74& 30000  &   17
         &  1.7$\times 10^{48}$     &  $\approx 10^{-6}$  \\
Z~And$^{5)}$    & 23/10/2008 & 1.5 &  5750 &11.5 & 0.05& 32000  & 3.1
         &  3.1$\times 10^{47}$     & 2.5$\times 10^{-6}$ \\
\hline\\[-6mm]
\end{tabular}
\end{center}
\scriptsize
$^{1)}$ in [$10^{37}$erg\,s$^{-1}$], ~$^{2)}$ in [$10^{60}$\cmt], ~
$^{3)}$ \cite{sk+14}, ~$^{4)}$ this paper, $^{5)}$ \cite{tarsko12}
\normalsize
\end{table}
\subsection{Symbiotic star Z~And}

Z~And is considered as a prototype of the class of symbiotic
stars. The donor is a late-type M4.5\,III giant, and the 
accretor is a WD on the 758-day orbit \citep[e.g.][]{nv89}. 
More than 120 years of its monitoring (first records from 1887) 
demonstrated an eruptive character of its LC. 
From 2000 September, Z~And started a series of outbursts with 
the main optical maxima in 2000 December, 2006 July, 2009 
December and 2011 November that peaked between 8 and 8.5 
in $U$ with an amplitude of $\sim 2-3$\,mag. During the 2006 
outburst, highly collimated bipolar jets were detected for 
the first time \citep[][]{sp06}. Example of its SED is 
depicted in Fig.~\ref{fig:seds}. 

\section{Concluding remarks}

Modelling the optical SED of active symbiotic stars Z~And and 
BF~Cyg revealed the simultaneous presence of a strong stellar 
and nebular component of radiation (see Fig.~\ref{fig:seds} 
and Table~1). The warm stellar component is not capable of 
producing the observed nebular emission, which signals the 
presence of a hot ionizing source in these systems, which is 
not seen directly by the observer. \cite{sk05} interpreted 
this two-temperature-type of the spectrum by the presence of 
an edge-on disk around the accretor, the outer flared 
rim of which represents the warm WD pseudophotosphere, and 
the nebula is placed above/below the disk, being ionized by 
the central hot star. 

Evolution of the optical/IR continuum of CN V339~Del during 
its fireball stage and the following transition to a harder 
spectrum suggests a similar ionization structure as that 
which appears to be present around WDs in symbiotic binaries 
during their outbursts 
\citep[see Sect.~4.2.3 of][ in detail]{sk+14}. 

The corresponding ionization structure of both very different 
types of interacting binaries is sketched in Fig.~2. 
This suggests that in spite of their extreme different 
energetic output, the basic ionization structure, which 
develops during the Z~And-type of the outburst, is similar 
to that which follows the fireball stage of CN outburst. 
Such a similarity suggests also a common mechanism of the 
mass ejection by the WDs in these systems. 
According to \cite{cask12}, the biconical ionization structure 
can be formed as a consequence of the enhanced mass-loss rate 
from the rotating WD during outbursts of symbiotic binaries 
(see their Figs. 1 and 6). This idea should be further tested by 
applying the wind compression disk model of \cite{bjorkcass93} 
to a rotating WD in CVs with parameters that represent 
CN outbursts. 
%
%-------------------------- Fig. 2 ------------------------
%
\begin{figure*}[!t]
%\centering
\begin{center}
%\resizebox{\hsize}{!}{\includegraphics[angle=-90]{skopal_f2.eps}}
\resizebox{12.5cm}{!}{\includegraphics[angle=-90]{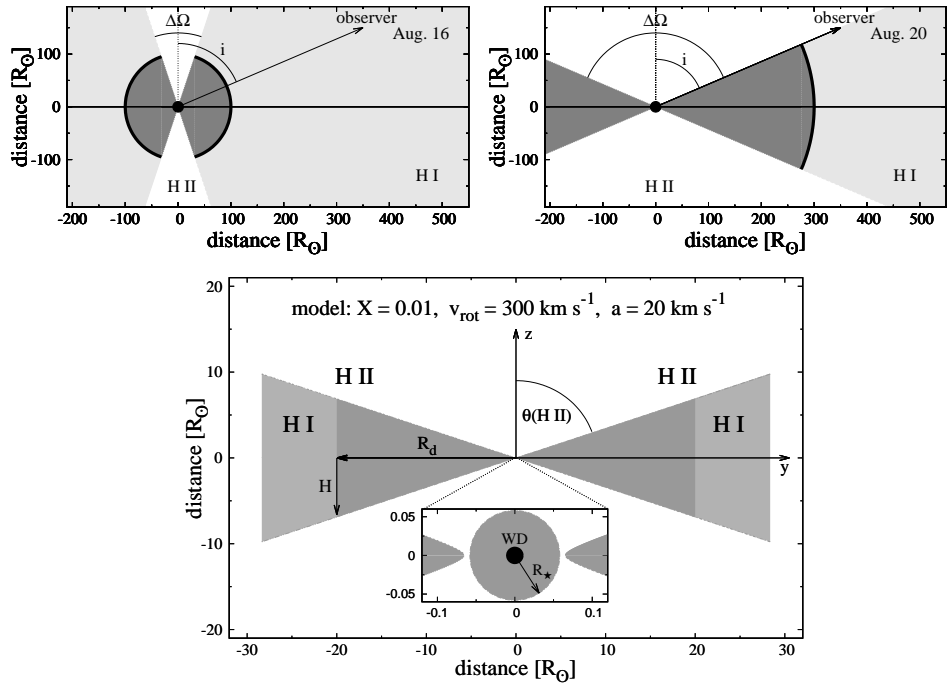}}
\end{center}
\vspace*{-2mm}
\caption[]{Top panels show a sketch of the ionization structure of 
CN V339~Del as seen on a cut perpendicular to the orbital plane 
containing the burning WD (the black circle). 
The WD pseudophotosphere is represented by the heavy solid line 
\citep[adapted from][]{sk+14}. Bottom panel shows a model of the
ionization structure of the hot components in symbiotic binaries 
during active phases \citep[adapted from][]{cask12}. 
          }
\label{fig:ion}
\vspace*{-2mm}
\end{figure*} 

\acknowledgements This research has been supported by a grant 
of the Slovak Academy of Sciences VEGA No.~2/0002/13. 
%
%\bibliography{editor}  % For BibTex
%
% For non-BibTex:

\end{document}